\documentclass[aps,prl,twocolumn,groupedaddress,showpacs,floatfix,letterpaper]{revtex4-1}%

\usepackage{graphicx}
\usepackage{bm}
\usepackage{latexsym}
\usepackage{epsf}
\usepackage{rotating}
\usepackage{epsfig,graphics,rotate,color}
\usepackage{wrapfig}
\usepackage{amssymb}
\usepackage{amsmath}
\usepackage{amsfonts}
\usepackage{siunitx}
\usepackage{array,hhline,dcolumn}%
\usepackage[normalem]{ulem}
\usepackage{color}
\usepackage{color}
\usepackage{hyperref}
\usepackage{graphicx}
\usepackage{mwe}
\begin{document}

\title{First Measurement of Monoenergetic Muon Neutrino Charged Current Interactions}

\date{\today}

\author{
        A.~A. Aguilar-Arevalo$^{13}$, 
        B.~C.~Brown$^{6}$, L.~Bugel$^{12}$,
	G.~Cheng$^{5}$, E.~D.~Church$^{20}$, J.~M.~Conrad$^{12}$, R.~L.~Cooper$^{10, 16}$,
	R.~Dharmapalan$^{1}$, 
	Z.~Djurcic$^{2}$, D.~A.~Finley$^{6}$, R.~S.~Fitzpatrick$^{14,*}$, R.~Ford$^{6}$,
        F.~G.~Garcia$^{6}$, G.~T.~Garvey$^{10}$, 
        J.~Grange$^{2,\dagger}$,
        W.~Huelsnitz$^{10}$, C.~Ignarra$^{12}$, R.~Imlay$^{11}$,
        R.~A. ~Johnson$^{3}$, J.~R.~Jordan$^{14,\ddagger}$, G.~Karagiorgi$^{5}$, T.~Katori$^{17}$,
        T.~Kobilarcik$^{6}$, 
        W.~C.~Louis$^{10}$, K. Mahn$^{5, 15}$, C.~Mariani$^{19}$, W.~Marsh$^{6}$,
        G.~B.~Mills$^{10}$,
	J.~Mirabal$^{10}$,
        C.~D.~Moore$^{6}$, J.~Mousseau$^{14}$, 
        P.~Nienaber$^{18}$, 
        B.~Osmanov$^{7}$, Z.~Pavlovic$^{10}$, D.~Perevalov$^{6}$,
         H.~Ray$^{7}$, B.~P.~Roe$^{14}$,
        A.~D.~Russell$^{6}$, 
	M.~H.~Shaevitz$^{5}$, 
        J.~Spitz$^{14,\S}$, I.~Stancu$^{1}$, 
        R.~Tayloe$^{9}$, R.~T.~Thornton$^{10}$, R.~G.~Van~de~Water$^{10}$, M.~O.~Wascko$^{8}$, 
        D.~H.~White$^{10}$, D.~A.~Wickremasinghe$^{3}$, G.~P.~Zeller$^{6}$,
        E.~D.~Zimmerman$^{4}$ \\
\smallskip
(MiniBooNE Collaboration)
\smallskip
}
\smallskip
\smallskip
\affiliation{
\mbox{$^1$University of Alabama, Tuscaloosa, Alabama 35487, USA} \\
\mbox{$^2$Argonne National Laboratory, Argonne, Illinois 60439, USA} \\
\mbox{$^3$University of Cincinnati, Cincinnati, Ohio 45221, USA}\\
\mbox{$^4$University of Colorado, Boulder, Colorado 80309, USA} \\
\mbox{$^5$Columbia University, New York, New York 10027, USA} \\
\mbox{$^6$Fermi National Accelerator Laboratory, Batavia, Illinois 60510, USA} \\
\mbox{$^7$University of Florida, Gainesville, Florida 32611, USA} \\
\mbox{$^8$Imperial College London, London SW7 2AZ, United Kingdom}\\
\mbox{$^9$Indiana University, Bloomington, Indiana 47405, USA}\\
\mbox{$^{10}$Los Alamos National Laboratory, Los Alamos, New Mexico 87545, USA} \\
\mbox{$^{11}$Louisiana State University, Baton Rouge, Louisiana 70803, USA} \\
\mbox{$^{12}$Massachusetts Institute of Technology, Cambridge, Massachusetts 02139, USA} \\
\mbox{$^{13}$Instituto de Ciencias Nucleares, Universidad Nacional Autónoma de M\'exico, D.F. 04510, Mexico} \\
\mbox{$^{14}$University of Michigan, Ann Arbor, Michigan 48109, USA} \\
\mbox{$^{15}$Michigan State University, East Lansing, Michigan 48824, USA} \\
\mbox{$^{16}$New Mexico State University, Las Cruces, New Mexico 88003, USA} \\
\mbox{$^{17}$Queen Mary University of London, London E1 4NS, United Kingdom} \\
\mbox{$^{18}$Saint Mary's University of Minnesota, Winona, Minnesota 55987, USA} \\
\mbox{$^{19}$Center for Neutrino Physics, Virginia Tech, Blacksburg, Virginia 24061, USA}\\
\mbox{$^{20}$Yale University, New Haven, Connecticut 06520, USA}\\
}

\begin{abstract}
We report the first measurement of monoenergetic muon neutrino charged current interactions. MiniBooNE has isolated 236~MeV muon neutrino events originating from charged kaon decay at rest ($K^+ \rightarrow \mu^+ \nu_\mu$) at the NuMI beamline absorber. These signal $\nu_\mu$-carbon events are distinguished from primarily pion decay in flight $\nu_\mu$ and $\overline{\nu}_\mu$ backgrounds produced at the target station and decay pipe using their arrival time and reconstructed muon energy. The significance of the signal observation is at the 3.9$\sigma$ level. The muon kinetic energy, neutrino-nucleus energy transfer ($\omega=E_\nu-E_\mu$), and total cross section for these events are extracted. This result is the first known-energy, weak-interaction-only probe of the nucleus to yield a measurement of $\omega$ using neutrinos, a quantity thus far only accessible through electron scattering.    
\end{abstract}
\maketitle

A charged kaon decays to a muon and a muon neutrino ($K^+ \rightarrow \mu^+ \nu_\mu$) 63.6\% of the time~\cite{pdg}. In the case that the kaon is at rest when it decays, the emitted muon neutrino is monoenergetic at 236~MeV. The kaon decay at rest (KDAR) neutrino has been identified as a gateway to a number of physics measurements, including searches for high-$\Delta m^2$ oscillations~\cite{kdar1,kdar3} and as a standard candle for studying the neutrino-nucleus interaction, energy reconstruction, and cross sections in the hundreds of MeV energy region~\cite{kdar2}. There are other ideas for using this neutrino, including as a source to make a precision measurement of the strange quark contribution to the nucleon spin ($\Delta s$)~\cite{kdar2} and as a possible signature of dark matter annihilation in the Sun~\cite{kumar,kumar2}. Despite the importance of the KDAR neutrino, it has never been isolated and identified.

In the charged current (CC) interaction of a 236~MeV $\nu_\mu$ (e.g.\ $\nu_\mu  \mathrm{^{12}C} \rightarrow \mu^- X$), the muon kinetic energy ($T_\mu$) and closely related neutrino-nucleus energy transfer ($\omega=E_\nu-E_\mu$) distributions are of particular interest for benchmarking neutrino interaction models and generators, which report widely varying predictions for kinematics at these transition-region energies~\cite{crpa,martini1,martini2,akbar,nuance,nuwro,genie,footnote}. Traditionally, experiments are only sensitive, at best, to total visible hadronic energy since invisible neutrons and model-dependent nucleon removal energy corrections prevent the complete reconstruction of energy transfer~\cite{minerva_q0}. The measurements reported here, therefore, provide a unique look at muon kinematics and the relationship to neutrino energy in the few hundreds of MeV range, highly relevant for both elucidating the neutrino-nucleus interaction and performing low energy precision oscillation measurements at short~\cite{miniboone_detector,microboone_detector,sbn} and long baselines~\cite{t2knim}.  

\begin{figure}[h]
\centering
\includegraphics[width=.48\textwidth]{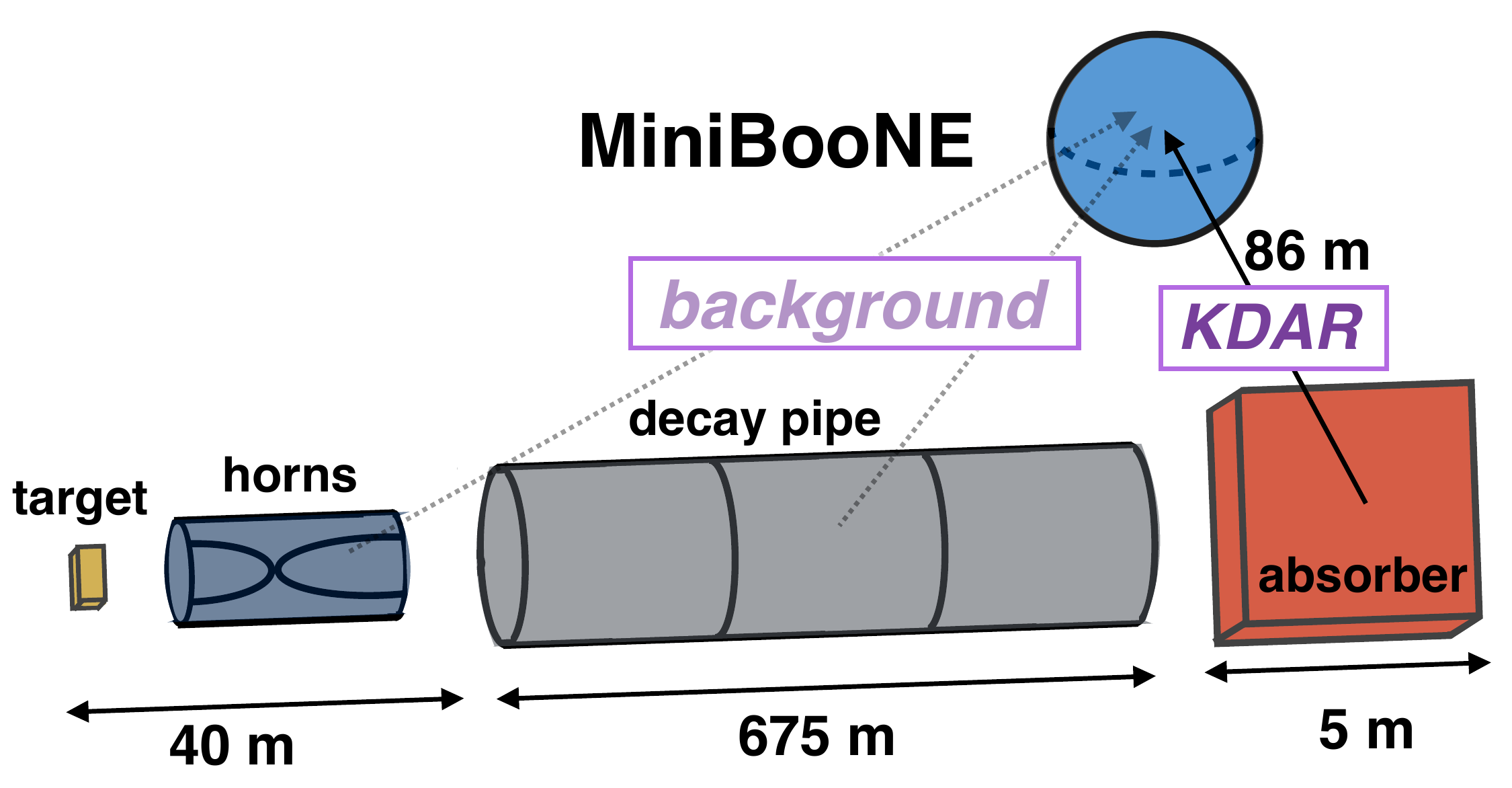} 
\vspace{-.5cm}
\caption{The NuMI beamline and the various sources of neutrinos that reach MiniBooNE (dashed lines). The signal KDAR neutrinos (solid line) originate mainly from the absorber.}
\label{fig0}
\end{figure}

The MiniBooNE detector uses 445~tons (fiducial volume) of mineral oil and 1280~photomultiplier tubes (PMTs), with an additional 240 PMTs instrumenting a veto region, to identify neutrino events originating from the Booster Neutrino Beamline (BNB) and Neutrinos at the Main Injector (NuMI) neutrino sources. The experiment has reported numerous oscillation and cross section measurements and new physics searches since data taking began in 2002~\cite{miniboone_detector}. For this analysis, we consider the charge and time data of PMT hits collected during the NuMI beam spill.  NuMI provides an intense source of KDAR neutrinos at MiniBooNE in a somewhat indirect way. The 96~cm, 2.0~interaction length NuMI target allows about 1/6 of the primary proton (120 GeV) power to pass through to the beam absorber~\cite{numi_beamline}, 725~m downstream of the target and 86~m from the center of MiniBooNE. The aluminum-core absorber, surrounded by concrete and steel, is nominally meant to stop the remnant hadrons, electrons, muons, and gammas that reach the end of the decay pipe. Interactions of primary protons with the absorber provide about 84\% of the total KDAR neutrinos from NuMI that reach MiniBooNE. Predictions from {\sc fluka}~\cite{flugg1,flugg2}, {\sc mars}~\cite{mars}, and {\sc geant4}~\cite{geant4} for kaon production at the absorber vary significantly, from 0.06--0.12~KDAR~$\nu_\mu$/proton on target. The background to the KDAR signal, $\nu_\mu$ and $\overline{\nu}_\mu$ CC events which produce a muon in the  0--115~MeV range, originates mainly from pion and kaon decay in flight near the target station and in the upstream-most decay pipe. The non-KDAR $\nu_\mu$ and $\overline{\nu}_\mu$ flux from the absorber, dominated by decay-in-flight kaons ($K_{\mu3}$ and $K_{\mu2}$) with a comparatively small charged pion component, is expected to contribute at the few-percent level based on a {\sc geant4} simulation of the beamline.   Figure ~\ref{fig0} shows a schematic of the NuMI beamline and its relationship to MiniBooNE. 

The KDAR event rate at MiniBooNE is expected to be similar in both NuMI's low-energy neutrino and antineutrino modes, since KDAR production from the absorber is not dependent on the polarization of the horns. However, the background $\nu_\mu$ and $\overline{\nu}_\mu$ event rate is predicted to be about 30\% lower in the antineutrino mode. We use data taken in this configuration from 2009--2011, corresponding to $2.62\times 10^{20}$~protons on the NuMI target.

The focus of this analysis is on reconstructing KDAR-like low energy $\nu_\mu$ CC events. A simple detector observable, \texttt{PMThits$_{\texttt{5ns}}$}, defined as the number of PMT hits multiplied by the fraction of light detected in the first 5~ns after correcting for vertex position, is used to reconstruct $T_\mu$ in selected events featuring \textbf{(1)} an electron from muon decay, noting that about 7.8\% of $\mu^-$ capture on nuclei~\cite{grange_thesis}, \textbf{(2)} a lack of veto activity, and \textbf{(3)} a reconstructed distance between the end point of the primary track and the muon decay vertex of $<$~150~cm. This detector observable is meant to isolate the muon via its characteristic prompt \v{C}erenkov light, as compared to the delayed scintillation-only light ($\tau=18$~ns) from the below-threshold hadronic part of the interaction. According to the {\sc nuwro} neutrino event generator~\cite{nuwro}, only 14\% of muons created in 236~MeV $\nu_\mu$ CC events are expected to be produced with energy less than 39~MeV, the \v{C}erenkov threshold for muons in MiniBooNE mineral oil. KDAR-induced muons are expected to populate a ``signal region", defined as 0--120~\texttt{PMThits$_{\texttt{5ns}}$} and representing $T_\mu$ in the range 0--115~MeV. Because of the kinematics of 236~MeV $\nu_\mu$ CC events, no signal is expected elsewhere, which is considered the ``background-only region" ($>$120~\texttt{PMThits$_{\texttt{5ns}}$}). Although the signal muon energy range considered for this measurement is lower than past MiniBooNE cross section analyses featuring $\nu_\mu / \overline{\nu}_\mu$~\cite{muons0,muons00,muons1,muons11,muons3,muons4,muons5}, the energy and timing distributions of MiniBooNE's vast calibration sample of 0--53~MeV electrons from muon decay provide a strong benchmark for understanding the detector's response to low energy muons in terms of both scintillation and \v{C}erenkov light. Further, a scintillator ``calibration cube" in the MiniBooNE volume at 31~cm depth, used to form a very pure sample of tagged $95\pm4$~MeV cosmic ray muons, shows excellent agreement between data and Monte Carlo simulations in terms of timing, \v{C}erenkov angle, and energy reconstruction~\cite{miniboone_detector}. The energy resolution for 95~MeV muons is measured to be 12\%; a detailed detector simulation agrees and predicts that the muon energy resolution in low energy $\nu_\mu$ CC events drops gradually to about 25\% for 50~MeV muons. The detection efficiency for KDAR $\nu_\mu$ CC events is $>$~50\% for events containing muons with energy $>$~50~MeV.

It is challenging to isolate the KDAR neutrino signal in MiniBooNE among the significant backgrounds. Even after optimizing event selection cuts and reconstruction, the signal-to-background ratio in the signal region is expected to only be $\sim$1:1. Along with the difficulty in identifying KDAR events based on neutrino energy, another issue is that reconstructing them as coming from the absorber is not possible because the muon and neutrino directions are poorly correlated at these low energies. Simply convolving a flux prediction with a neutrino cross section to form a background rate prediction is also not feasible. Although a reliable background flux prediction is available~\cite{leo_thesis}, the rate and kinematics of such events in MiniBooNE are also determined by the relevant total and differential $\nu_\mu$ and $\overline{\nu}_\mu$ cross sections for neutrino energies in the hundreds of MeV region. The rapid turn on of the $\nu_\mu$ CC cross section above the mass of the muon and almost complete lack of data below 400~MeV~\cite{anl,lsnd_xsec} would make any kind of background prediction at these energies arbitrary and highly uncertain.

In order to mitigate the issues associated with the background prediction near KDAR energies, we use a timing-based \emph{in situ} background measurement technique which relies on the fact that KDAR $\nu_\mu$ CC events from the absorber arrive at MiniBooNE $\sim$200~ns after background $\nu_\mu$ and $\overline{\nu}_\mu$ CC events originating from the target station and decay pipe. The background neutrinos simply take a more direct route to MiniBooNE as compared to their signal counterparts from the absorber: the distance from the target to the absorber plus the distance from the absorber to MiniBooNE is $725+86=811$~m while the distance from the target to MiniBooNE is 749~m. Although the beam window is $\sim$9~$\mu$s, this timing difference provides a ``background-enhanced" period at the beginning of the window, where background $\nu_\mu$ and $\overline{\nu}_\mu$ CC events are expected to dominate, and a ``signal-enhanced" period at the end of the window, where signal KDAR $\nu_\mu$ CC events from the absorber dominate. Considering the neutrino event timing resolution and the timing uncertainties due to various sources, we define the first and last 600~ns of the beam window as background and signal enhanced, respectively. The inset of Fig. 2 shows the relative event rate in the enhanced regions compared to a high-statistics region in which signal and background remain constant (referred to as ``normal time'' and discussed later in detail). Most notably, there is a 2.4$\sigma$ (2.1$\sigma$) excess (deficit) of KDAR-like events (0--120~\texttt{PMThits$_{\texttt{5ns}}$}) at late (early) times.

In the absence of a reliable background prediction, we employ a template-based analysis which tests the consistency of various candidate KDAR signal $T_\mu$ distributions with data. We consider a broad and well-defined set of possible $T_\mu$ signal shapes and determine how well each matches the data. This procedure can be thought of as the reverse of the usual differential cross section measurement extraction. Instead of starting from a detector observable and turning it into a measure of $T_\mu$, for example, we start with a candidate ``true" $T_\mu$ distribution and and map (or ``fold") it into a detector observable distribution in \texttt{PMThits$_{\texttt{5ns}}$}. The candidate true $T_\mu$ signal shapes are based on a beta distribution. This carefully chosen function, with only two parameters characterizing its shape, is meant to cover all physical and continuous shapes that the true KDAR-induced $T_\mu$ distribution can take, noting that we are not sensitive to few-MeV-scale resonance features (e.g. as predicted by continuous random phase approximation calculations~\cite{crpa}). The shape of the signal model ($T_\mu$ spectrum) is defined by two parameters, $a$ and $b$, according to the beta distribution:
$x^{a-1} (1-x)^{b-1}/B(a,b)$~, 
where $B(a,b) = \Gamma(a)\Gamma(b)/\Gamma(a+b)$ and $x=T_\mu/T_\mu^\mathrm{max}$. After correcting for detector efficiency, each candidate $T_\mu$  distribution is folded into the corresponding \texttt{PMThits$_{\texttt{5ns}}$}  distribution and compared to data as a function of time. The normalizations of signal and background are expected to change at early and late times, but the shapes of each stay nearly constant. 

\begin{figure}[t!]
\centering
\includegraphics[width=.49\textwidth]{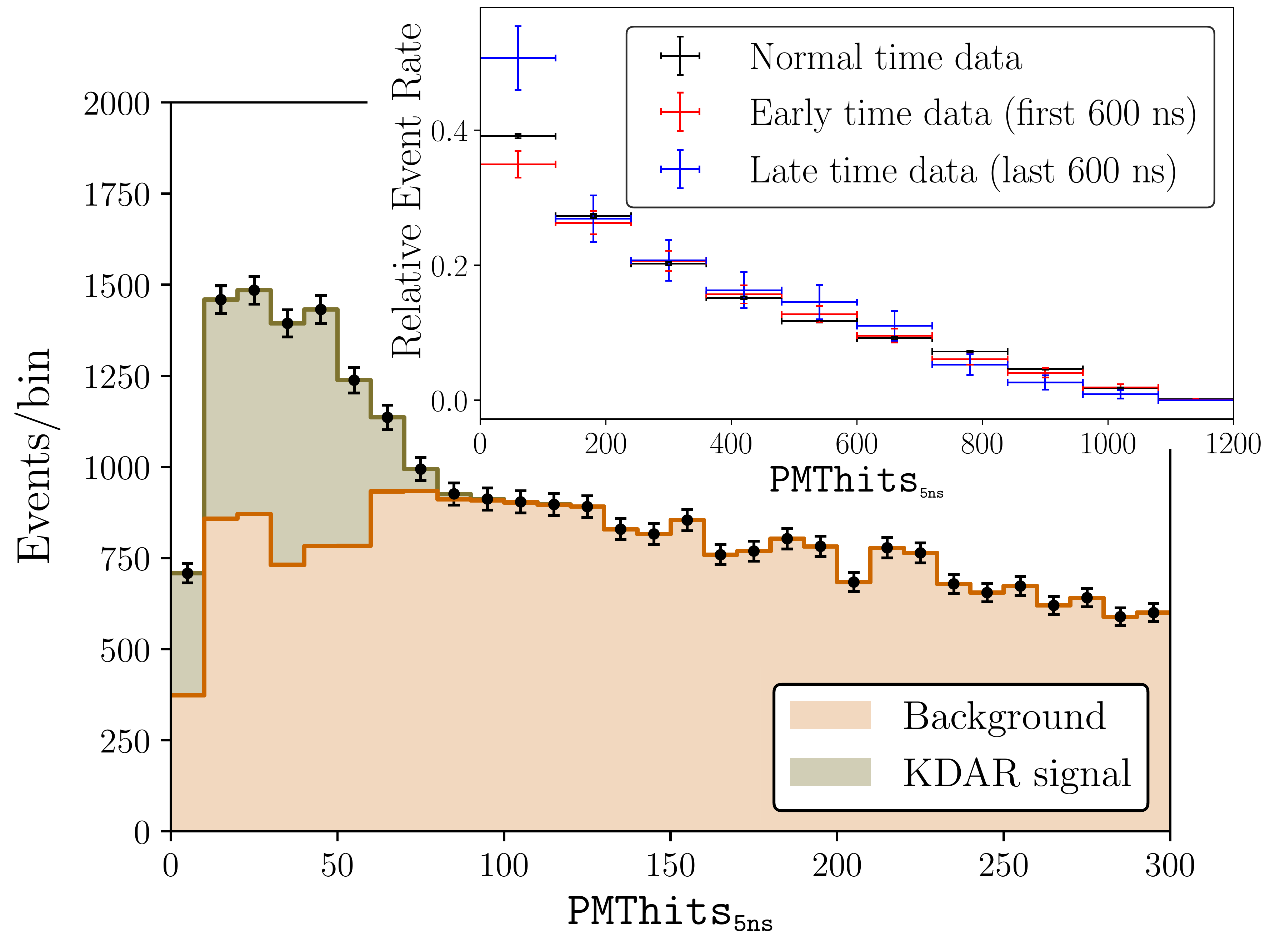}
\vspace{-.7cm}
\caption{The normal time data distribution (black points with error bars) with the best-fit signal template (green) stacked on the inferred background (orange). The inset shows the relative event rate for early time, late time, and normal time after normalizing the three distributions in the background-only region (\texttt{PMThits$_{\texttt{5ns}}$}$>$120). A deficit (excess) of KDAR-like events at early (late) times can be seen.}
\label{fig2}
\end{figure}

\begin{figure}[t!]
\begin{tabular}{@{\hspace{-.19cm}}c@{\hspace{-.08cm}}c}
\vspace{-.66cm} 
\includegraphics[scale=0.31]{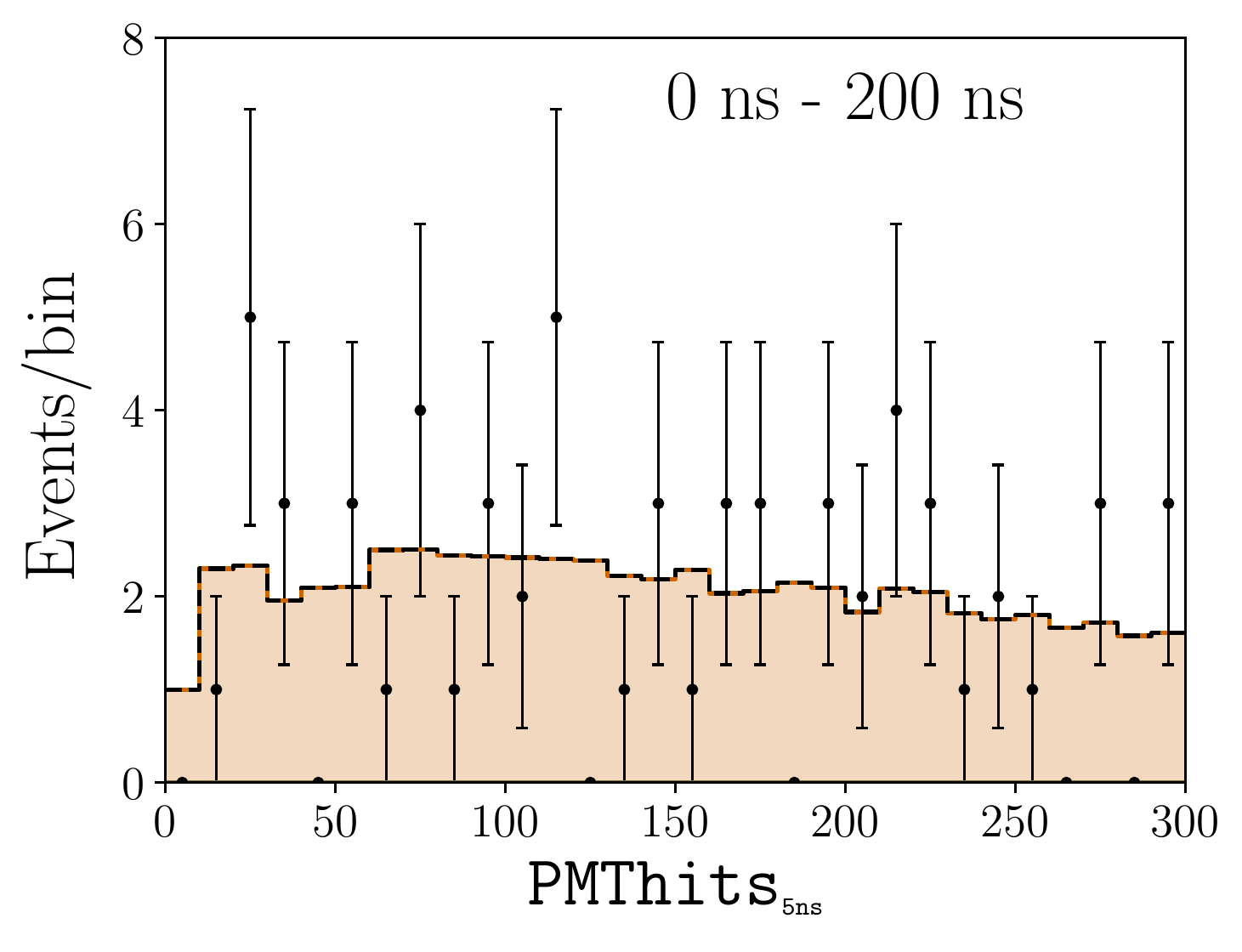} & 
\includegraphics[scale=0.31,trim={31 0 0 0},clip]{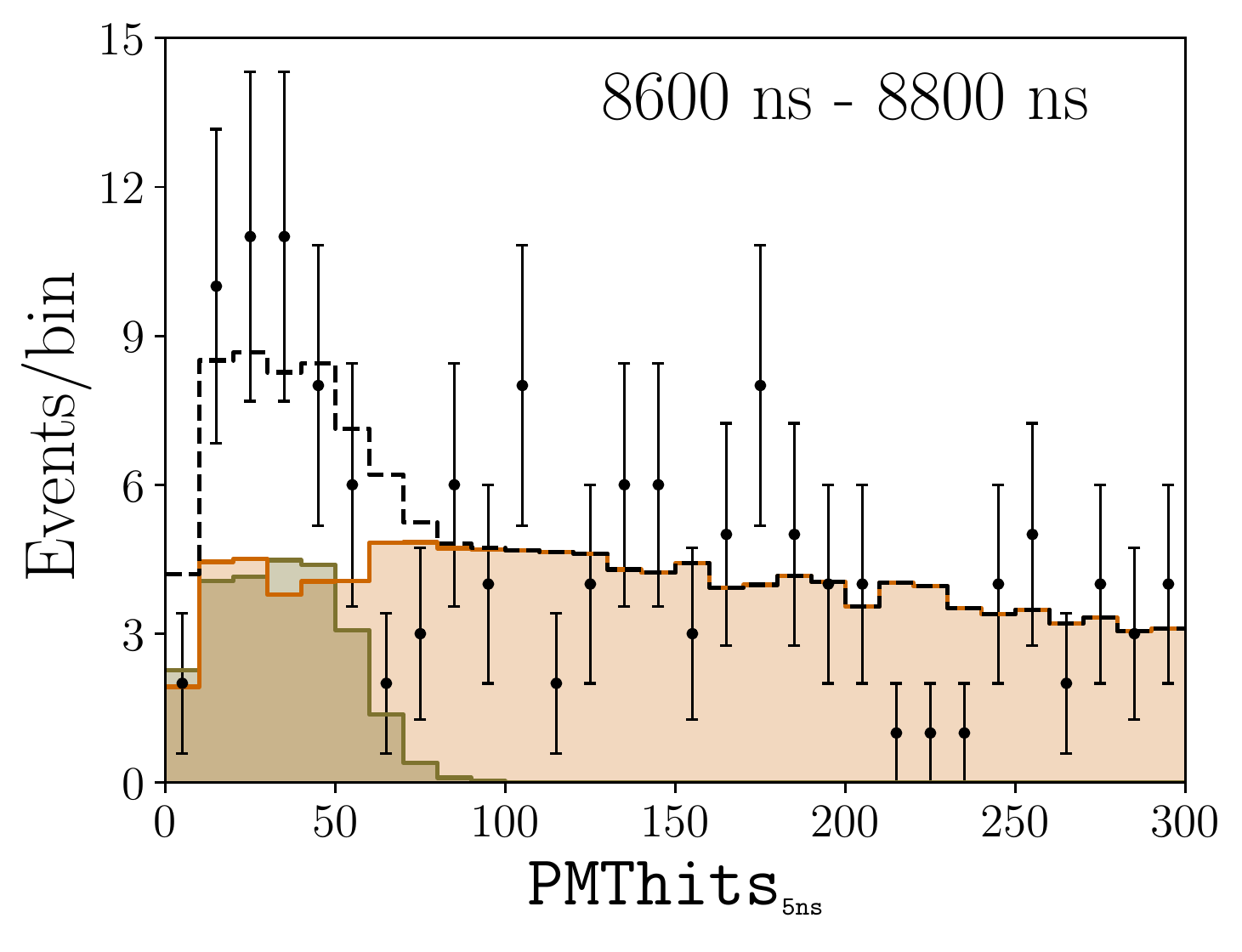}\\
\vspace{-.66cm} 
\includegraphics[scale=0.31]{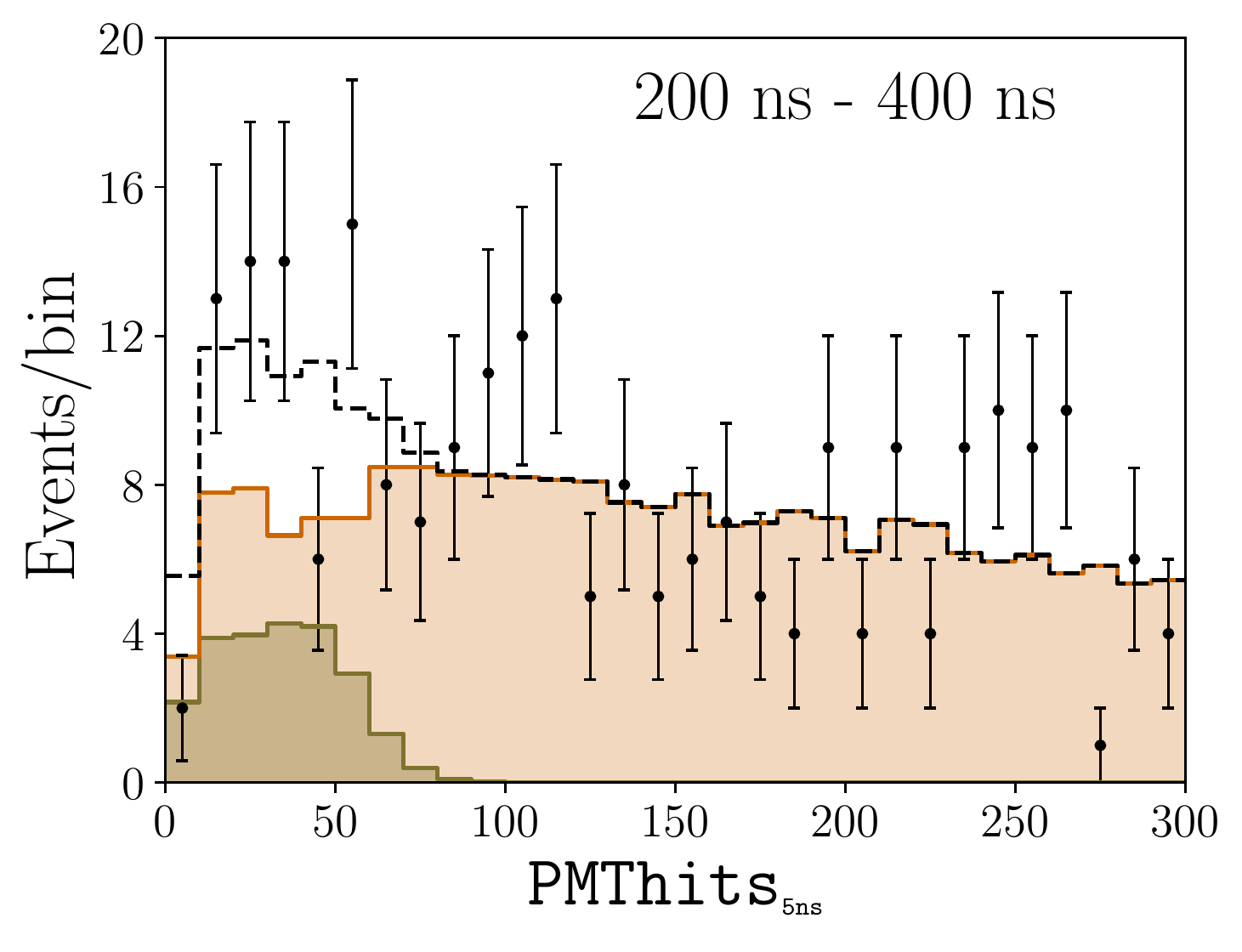} &
\includegraphics[scale=0.31,trim={31 0 0 0},clip]{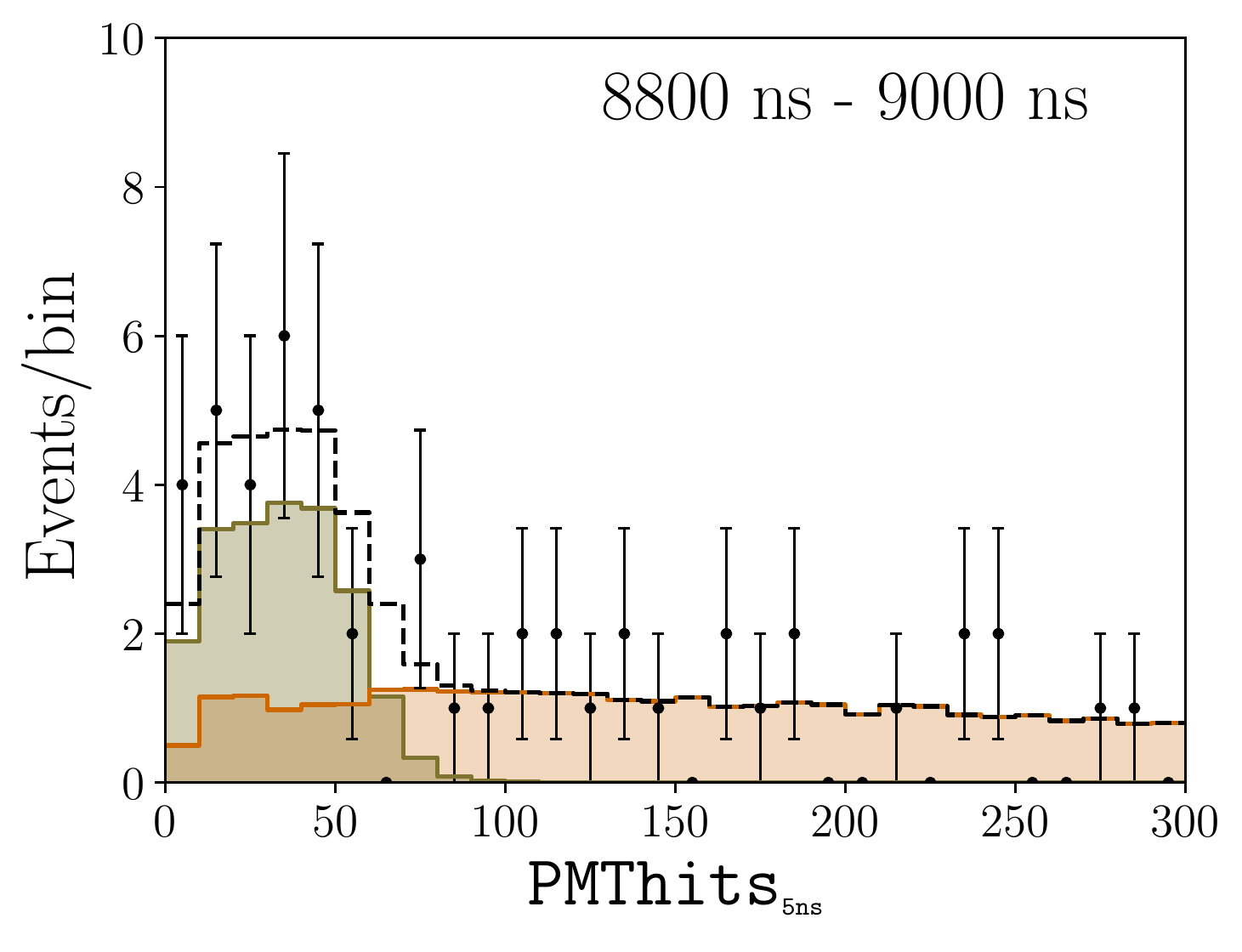}\\
\vspace{-.66cm}
\includegraphics[scale=0.31]{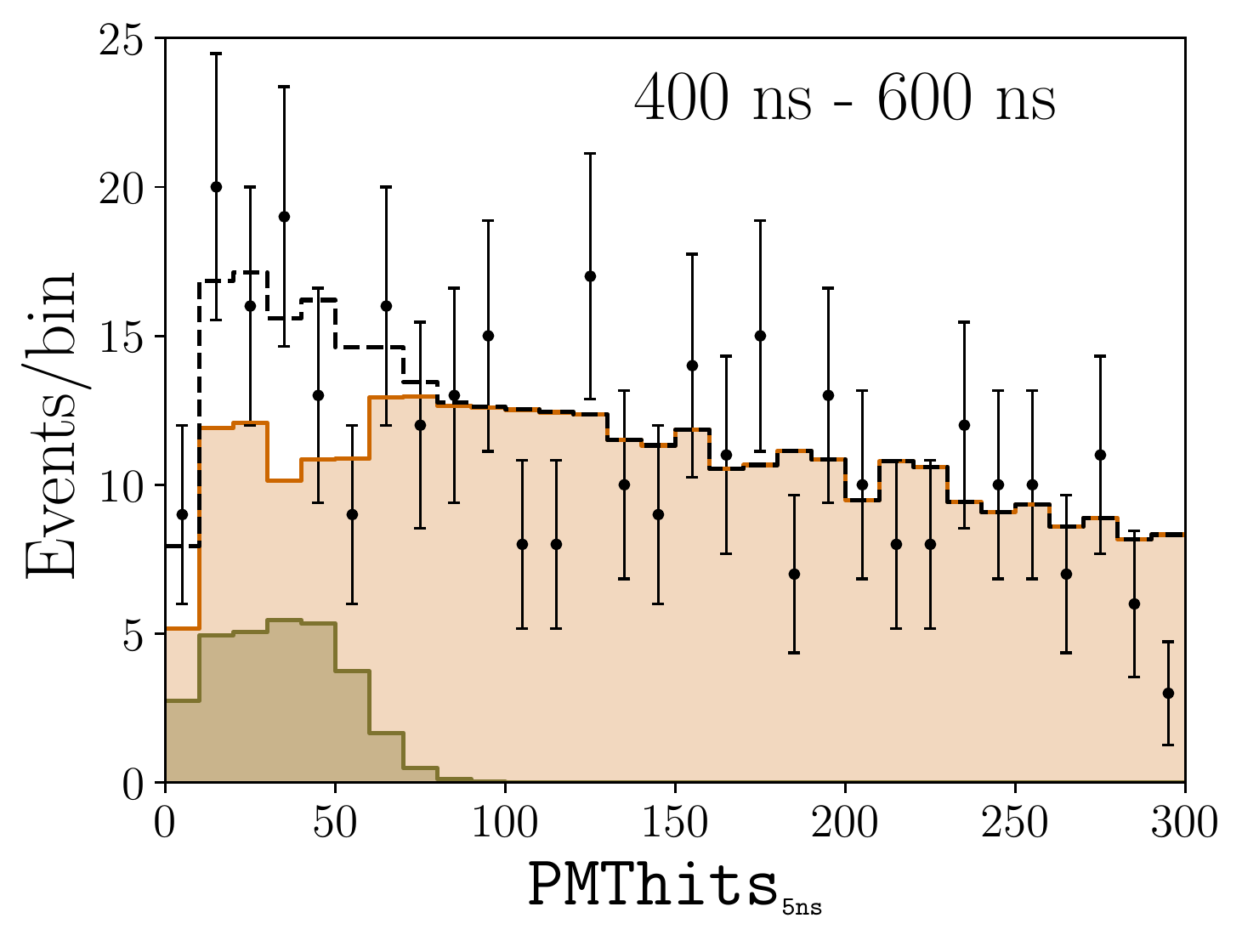} &
\includegraphics[scale=0.31,trim={31 0 0 0},clip]{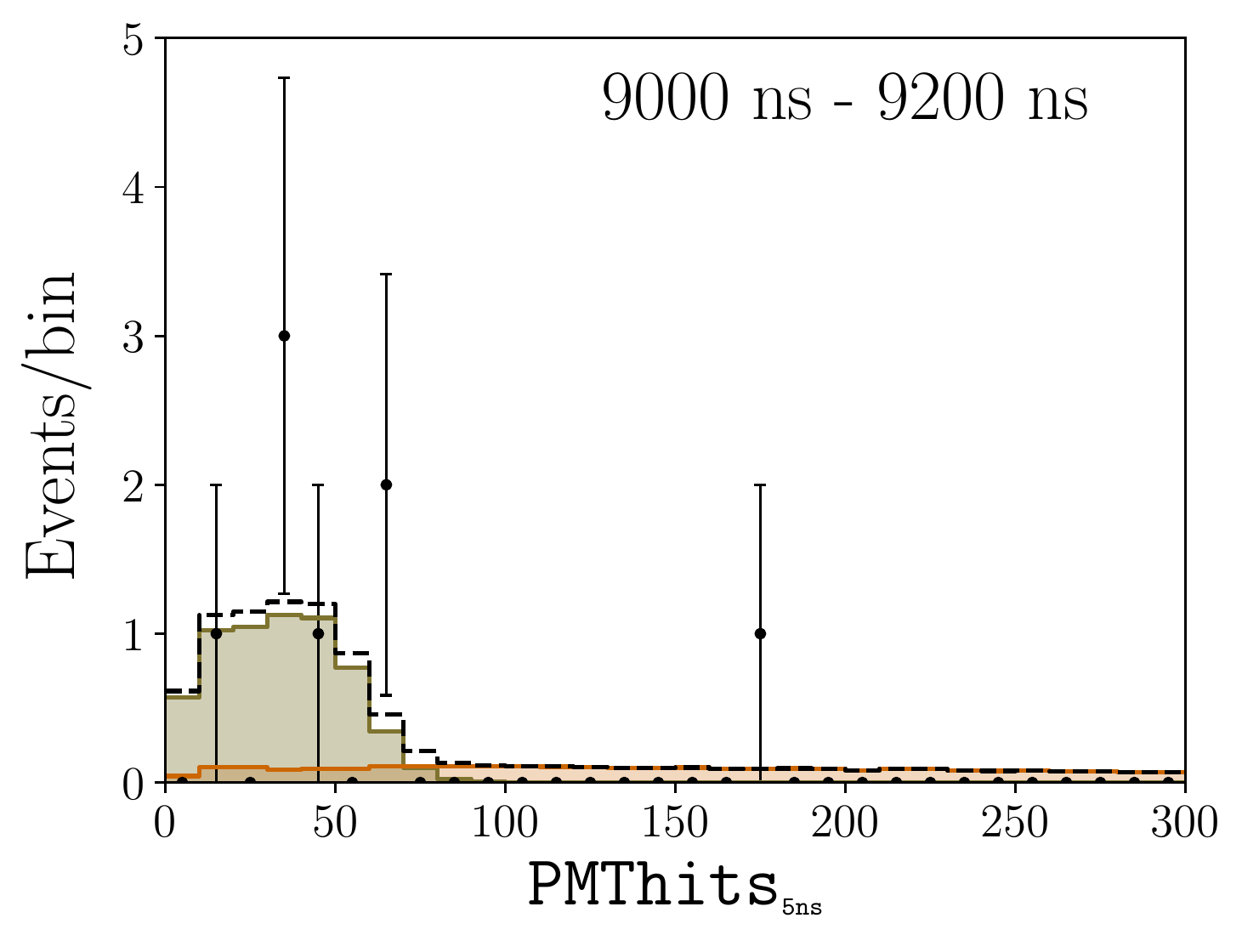}
  \end{tabular}
\vspace{0.3cm}
\caption{(Left) The three early-time, background-enhanced bins. (Right) The three late-time, signal-enhanced bins. The data (black solid line with stat-only error bar), best-fit signal (green), best-fit background (orange), and total signal+background (black dotted-line) distributions are shown.}
%\vspace{-.5cm}
\label{timebins}
\end{figure}

The analysis proceeds in four steps. \textbf{(1)} The data sample is broken up into seven time bins within the 9200~ns beam window: three early-time bins (200~ns each), one ``normal-time" (NT) bin (8000~ns), and three late-time bins (200~ns each). \textbf{(2)} Using the following procedure, signal and background templates in \texttt{PMThits$_{\texttt{5ns}}$} are formed using the high-statistics NT region distribution as a reference, where signal and background are expected to be constant. The candidate signal template is drawn from a large number of possible shapes and normalizations within reasonable physical limits. In the signal region (0--120~\texttt{PMThits$_{\texttt{5ns}}$}), the background template is defined such that the candidate signal plus background distribution is equal to the NT data. Figure~\ref{fig2} shows an example set of templates overlaid on data in NT. \textbf{(3)} In each of the three early-time and three late-time bins, the normalization of each background template is adjusted so that it is consistent with the number of events observed in each time bin's background-only region. Figure ~\ref{timebins} shows an example set of constant-shaped background templates for each of the early-time and late-time bins.
\textbf{(4)} A Poisson extended maximum likelihood $\chi^2$ statistic~\cite{pdg, cowan, roe} is formed from a comparison between the signal+background templates and data in the signal regions of the three early-time and three late-time bins. This treatment is studied later with Monte Carlo simulations. The procedure is then repeated for various combinations of candidate signal shapes, normalizations, and end points.
%\end{enumerate}

For a particular time bin ($i$), excluding the NT bin, the signal region data are distributed into 12~\texttt{PMThits$_{\texttt{5ns}}$} bins ($j$) from 0--120. A $\chi^2_i$ for a Poisson-distributed variable is then formed by comparing the data ($d_{i,j}$) and a prediction ($P_{i,j,\alpha}$) based on the signal model ($T_{j,\alpha}$) with signal normalization $\alpha$ plus the background ($B_{i,j}$) such that $P_{i,j,\alpha}=T_{j,\alpha}+B_{i,j}$:
\begin{align*}
\begin{split}
\chi^2_{i,\alpha} = 2 \sum_j \begin{cases} 
P_{i,j,\alpha} - d_{i,j} + d_{i,j} \ln(d_{i,j}/P_{i,j,\alpha})  & d_{i,j} > 0 \\
P_{i,j,\alpha} & d_{i,j} = 0~.
\end{cases}
\end{split}
\end{align*}
We then marginalize over the signal normalizations in each time bin to produce $\chi^2_i = \min\limits_{\alpha}(\chi^2_{i,\alpha})$. 


No KDAR signal events from the absorber are expected in the first 200~ns time bin. This time period, therefore, contains the expected \texttt{PMThits$_{\texttt{5ns}}$} shape of the background distribution in the signal region. In the first time bin, the measured ratio of data events in the 0--120~\texttt{PMThits$_{\texttt{5ns}}$} signal region (28) to total number of events (118) is compared to the equivalent ratio for the current candidate model's background prediction to form an uncertainty weighted pull term ($f_\mathrm{pull}$). This pull term penalizes candidate models that produce background templates inconsistent with the first time bin. Finally, the total $\chi^2$ for a particular model shape and normalization is given by $\chi^2 = \sum_i \chi^2_i + f_\mathrm{pull}$.

We test a set of physically allowed and reasonable models with the parameter sets $a \in [2.0, 8.0]$, $b \in [0.0, 6.0]$. Models with $0.0<a<2.0$ are considered unphysical and inconsistent with all predictions since they are initially concave down or do not go to zero at $T_\mu=0$~MeV. We also test a range of muon kinetic energy ``effective end points," $T_\mu^\mathrm{max}=95$--115~MeV. Although the separation energy in $^{12}$C is 17~MeV, corresponding to a $T_\mu$ end point of 112~MeV, we consider this range of effective end points for capturing the characteristic behavior of the distribution near threshold, limited by the coarse sensitivity of a two-parameter model. 

\begin{figure}[t]
\centering
\hspace{-.25cm}
\includegraphics[width=.49\textwidth]{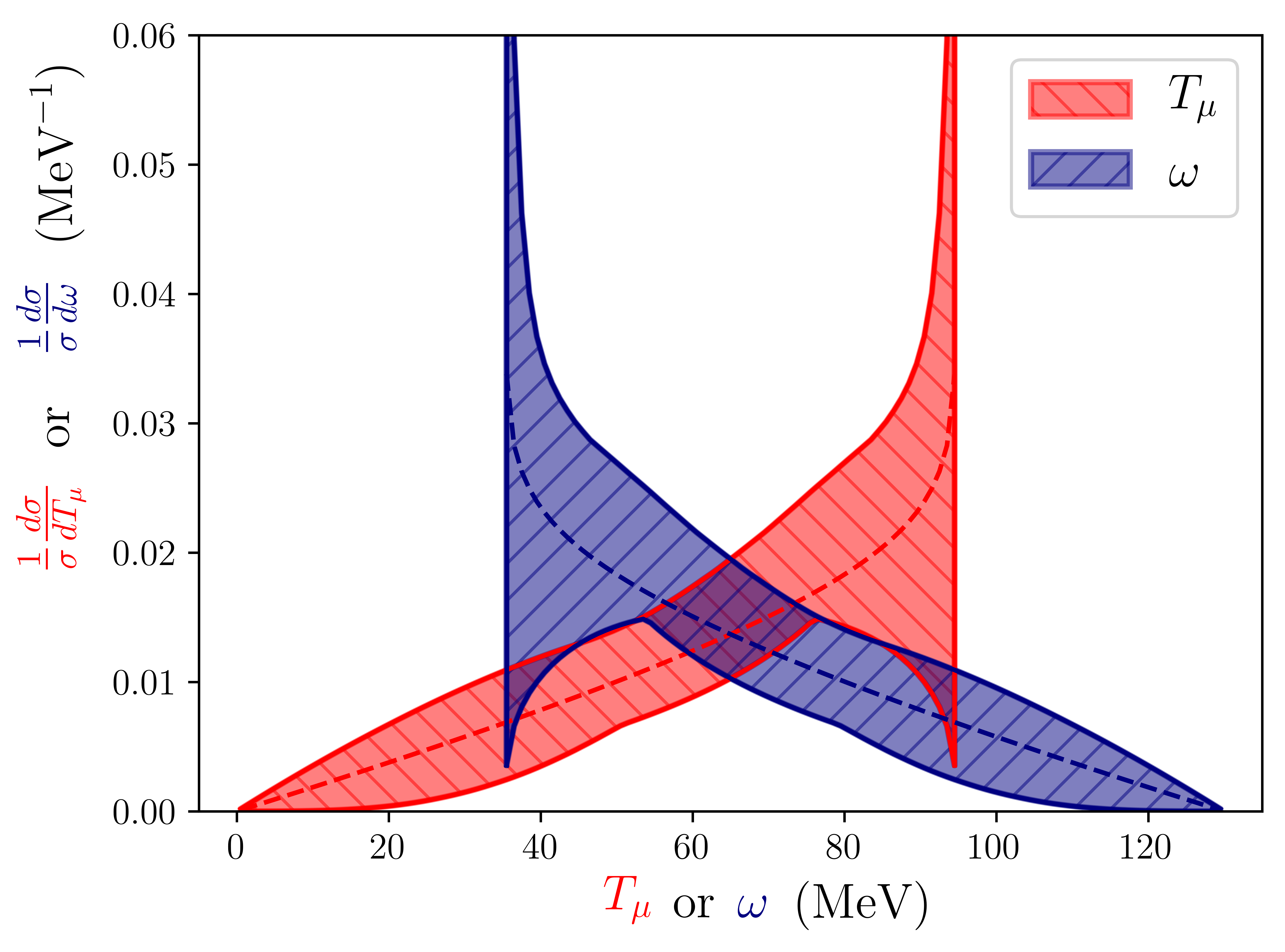}
\vspace{-.7cm}
\caption{The best fit $T_\mu$ (red-dashed) and $\omega$ (blue-dashed) spectra with shape-only 1$\sigma$ error bands, given a fixed end point of $T_\mu^\textrm{max}=95$~MeV. The distributions are fully correlated.}
\label{fig3}
\end{figure}

The best fit model parameters found are $a = 2.0$, $b = 0.88$, with a signal normalization of $3700\pm 1250$ events ($\chi^2_\mathrm{min} = 72.6$ with 64~degrees of freedom). The NT data and best fit signal and background distributions are shown in Fig.~\ref{fig2} and the corresponding results for each early- and late-time bin are shown in Fig.~\ref{timebins}. The extracted $T_\mu$ and $\omega=236~\mathrm{MeV}-m_\mu-T_\mu$ distributions with $1\sigma$ ($\chi^2_\mathrm{min}+2.3$) shape-only allowed bands are shown in Fig.~\ref{fig3}. The result is shown with $T_\mu^\mathrm{max} = 95$~MeV, representing the best fit effective end point, noting that $T_\mu^\mathrm{max}$ values up to the physical limit of 112~MeV are not strongly disfavored. A simulation with events distributed according to the best fit shape and data normalizations in each time bin confirms that the size of the 1$\sigma$ allowed region is reasonable, with 61\% (65\%) of best fit values falling in the 2 (3) parameter shape-only (rate+shape) contour. In the case that the end point is included as an additional shape parameter, we find that 66\% of best fit values fall in the three parameter shape+end point contour.

In addition to the parametrization-based results reported here, a data release~\cite{website} associated with this measurement has been made available which allows one to compare an \textit{arbitrary} $T_\mu$ or $\omega$ shape and end point prediction with the data. The program takes a prediction, folds it into the detector observable, corrects for detector efficiency, and then performs a comparison to data. This comparison is straightforward because, given the short background-enhanced and signal-enhanced timing windows, statistical uncertainties dominate. Although \num{49514}~events pass the selection cuts across all reconstructed energies, only 310 (115) events from the background- (signal-)enhanced period enter the KDAR $\nu_\mu$ signal region (0--120 \texttt{PMThits$_{\texttt{5ns}}$}). The fractional error contribution of systematics, including those associated with the optical model and detector response, across the full kinematic range is at the 1\%--2\% level. 

In order to determine the significance of the KDAR $\nu_\mu$ observation, we compare the best-fit result ($\chi^2_\mathrm{min} = 72.6$) to a zero-parameter, background-only hypothesis ($\chi^2_\mathrm{null}=113.8$). Simulated data, created by distributing events according to the background-only hypothesis and the data normalization in each time bin, are used to study the significance of this result. We find that the probability of obtaining a $\Delta \chi^2=\chi^2_\mathrm{null}-\chi^2_\mathrm{min}>41.2$ is about $1.1\times10^{-4}$, corresponding to 3.9$\sigma$ (two sided).

The main motivation for presenting a shape-only differential cross section measurement, rather than a normalized one, is that there is a large uncertainty on the kaon production at the absorber. However, we can report a coarse total cross section value after assigning a conservative 30\% uncertainty to the prediction by {\sc geant4} for the flux from the absorber (0.085~KDAR $\nu_\mu$/proton on target). We extract a total $\nu_\mu$ CC cross section at $E_\nu=236$~MeV of $\sigma=(2.7 \pm0.9 \pm 0.8)\times 10^{-39}$~cm$^2$/neutron. The first error represents the total uncertainty of the rate+shape measurement, and the second comes from the uncertainty on the initial KDAR flux. Adding these in quadrature yields $\sigma=(2.7\pm1.2)\times 10^{-39}$~cm$^2$/neutron. This can be compared to the {\sc nuwro} prediction of $\sigma=1.3\times 10^{-39}$~cm$^2$/neutron~\cite{nuwro}.

Building on the measurements presented here, the KDAR neutrino will be studied in great detail in the near future. MicroBooNE (102~m from the NuMI absorber)~\cite{microboone_detector}, for example, will be able to use its imaging capabilities to precisely study the hadronic part of the KDAR interaction, reconstruct the muon direction, and mitigate background via neutrino direction reconstruction. Scaling from the measurement reported here, we expect over 2000 KDAR events have already been collected by MicroBooNE, which continues taking data. In addition, the J-PARC Sterile Neutrino Search at the J-PARC Spallation Neutron Source (JSNS$^2$) will study KDAR muon neutrinos with excellent muon energy resolution and negligible background; \num{10000}--\num{20000} KDAR $\nu_\mu$ CC events per year are expected after JSNS$^2$ starts taking data in approximately one year~\cite{jsns2_tdr}. 

In summary, MiniBooNE has performed the first measurement of monoenergetic $\nu_\mu$ CC events. The 236~MeV KDAR neutrinos, originating at the NuMI absorber 86~m from MiniBooNE, are distinguished from background neutrinos created at the NuMI target station and decay pipe using muon energy reconstruction and timing. We have employed a somewhat unconventional analysis, which relies on a parametrized $T_\mu$ prediction and subsequent comparison to data, for extracting the result. This data-driven measurement does not rely on unfolding and is largely independent of both cross section and kinematic predictions from neutrino event generators and a flux determination. These results provide a standard candle benchmark, in terms of a variable historically unavailable to neutrino scattering experiments ($\omega$), for modeling the relationship between lepton kinematics and neutrino energy, elucidating the neutrino-nucleus to neutrino-nucleon transition region, and using the associated predictions to inform oscillation measurements at short and long baselines.   

We acknowledge Fermilab, the Department of Energy, the National Science Foundation, and Los Alamos National Laboratory for support of this experiment. Also, we thank Leonidas Aliaga, the MINERvA Collaboration, and Robert Hatcher for help understanding the NuMI flux and accelerator information. 

\vspace{12pt}

\noindent$^*$roryfitz@umich.edu\\
$^\dagger$grange@anl.gov\\
$^\ddagger$jrlowery@umich.edu\\
$^\S$spitzj@umich.edu

%Unused bibitems

%\bibitem{uboone_prop}
%H.~Chen \textit{et al.} [MicroBooNE Collaboration], ``Proposal for a New Experiment Using the Booster and NuMI Neutrino Beamlines: MicroBooNE",
%FERMILAB-PROPOSAL-0974 (2007). 
% 

\begin{thebibliography}{9}

\bibitem{pdg}
J. Beringer \textit{et al.}  (Particle Data Group), 
Phys. Rev. D \textbf{86} 010001 (2012).

\bibitem{kdar1}
J. Spitz, 
 Phys.\ Rev.\  D {\bf 85} 093020 (2012).
 
\bibitem{kdar3}
S. Axani, G. Collin, J.M. Conrad, M.H. Shaevitz, J. Spitz, T. Wongjirad, Phys. Rev. D \textbf{92} 092010 (2015).

\bibitem{kdar2}
J. Spitz, 
 Phys.\ Rev.\  D {\bf 89} 073007 (2014).

\bibitem{kumar}
C. Rott, S. In, J. Kumar, and D. Yaylali, J. of Cosmol. and Astropart. Phys. 11 (2015) 039.

\bibitem{kumar2}
C. Rott, S. In, J. Kumar, and D. Yaylali, arXiv:1710.03822.

\bibitem{crpa}
V. Pandey, N. Jachowicz, T. Van Cuyck, J. Ryckebusch, and M. Martini, 
Phys. Rev. C \textbf{92} 024606 (2015).

\bibitem{martini1}
M. Martini, M. Ericson, G. Chanfray, and J. Marteau,
Phys. Rev. C \textbf{80} 065501 (2009).

\bibitem{martini2}
M. Martini, M. Ericson, and G. Chanfray,
Phys. Rev. C \textbf{84} 055502 (2011).

\bibitem{akbar}
F. Akbar, M. Sajjad Athar, and S.K. Singh,
J. Phys. G \textbf{44} 125108 (2017).

\bibitem{nuance}
D. Casper,
Nucl. Phys. B, Proc. Suppl. \textbf{112} 161 (2002). 

\bibitem{nuwro}
C. Juszczak, 
Acta Phys. Pol. B \textbf{40} 2507 (2009);
T.~Golan, C. Juszczak, and J. T. Sobczyk,
Phys. Rev. C \textbf{86} 015505 (2012).

\bibitem{genie}
C. Andreopoulos \textit{et al.}, 
Nucl. Instrum. Methods Phys. Res., Sect. A \textbf{614} 87 (2010).

\bibitem{footnote}
We have compiled a number of predictions for the KDAR neutrino's outgoing muon kinetic energy in the data release associated with this measurement~\cite{website}.

\bibitem{website}
\url{https://www-boone.fnal.gov/for_physicists/data_release/kdar/}


\bibitem{minerva_q0}
P.A. Rodrigues \textit{et al.} (MINERvA Collaboration),
Phys. Rev. Lett. \textbf{116} 071802 (2016).

\bibitem{miniboone_detector}
A.A. Aguilar-Arevalo \textit{et al.} (MiniBooNE Collaboration),
 Nucl. Instrum. Methods Phys. Res., Sect. A \textbf{599} 28 (2009).

\bibitem{microboone_detector}
R. Acciarri \textit{et al.} (MicroBooNE Collaboration),
J. Instrum. \textbf{12} P02017 (2017). 

\bibitem{sbn}
R. Acciarri \textit{et al.},
arXiv:1503.01520.


\bibitem{t2knim}
K. Abe \textit{et al.} [T2K Collaboration],
Nucl. Instrum. Methods Phys. Res., Sect. A \textbf{659} 106 (2011).

\bibitem{numi_beamline}
P. Adamson \textit{et al.}, 
Nucl. Instrum. Methods Phys. Res., Sect. A \textbf{806} 279 (2016).

\bibitem{flugg1}
A. Ferrari, P.R. Sala, A. Fassò and J. Ranft, 
Reports No. CERN-2005-010, SLAC-R-773, INFN-TC-05-11.

\bibitem{flugg2}
T.T. Böhlen, F. Cerutti, M.P.W. Chin, A. Fassò, A. Ferrari, P.G. Ortega, A. Mairana, P.R. Sala, G. Smirnov, and V. Vlachoudis, Nucl. Data Sheets \textbf{120}, 211 (2014).

\bibitem{mars}
N.V.~Mokhov, Report No. FERMILAB-FN-628, 1995;
O.E.~Krivosheev and N.V. Mokhov, Report No. Fermilab-Conf-00/181, 2000;
O.E.~Krivosheev and N.V. Mokhov, Report No. Fermilab-Conf-03/053, 2003;
N.V.~Mokhov, K.K.~Gudima, C.C.~James \textit{et al.},  Report No. Fermilab-Conf-04/053, 2004.

\bibitem{geant4}
S. Agostinelli \textit{et al.}, 
Nucl. Instrum. Methods Phys. Res., Sect. A \textbf{506} 250 (2003).

\bibitem{grange_thesis}
J. Grange, Ph.D. thesis, University of Florida, 2013.

\bibitem{muons0}
A.A. Aguilar-Arevalo \textit{et al.} (MiniBooNE Collaboration),
Phys. Rev. Lett. \textbf{100} 032301 (2008).

\bibitem{muons00}
A.A. Aguilar-Arevalo \textit{et al.} (MiniBooNE Collaboration),
Phys. Rev. Lett. \textbf{103} 061802 (2009).

\bibitem{muons1}
A.A. Aguilar-Arevalo \textit{et al.} (MiniBooNE Collaboration),
Phys. Rev. Lett. \textbf{103} 081801 (2009).

\bibitem{muons11}
A.A. Aguilar-Arevalo \textit{et al.} (MiniBooNE Collaboration),
Phys. Rev. D \textbf{81} 092005 (2010).

\bibitem{muons3}
A.A. Aguilar-Arevalo \textit{et al.} (MiniBooNE Collaboration),
Phys. Rev. D \textbf{83} 052007 (2011).

\bibitem{muons4}
A.A. Aguilar-Arevalo \textit{et al.} (MiniBooNE Collaboration),
Phys. Rev. D \textbf{83} 052009 (2011).

\bibitem{muons5}
A.A. Aguilar-Arevalo \textit{et al.} (MiniBooNE Collaboration),
Phys. Rev. D \textbf{88} 032001 (2013).


\bibitem{leo_thesis}
L. Aliaga Soplin,  Ph.D. thesis, College of William and Mary (2016).

\bibitem{anl}
S.J. Barish \textit{et al.},
Phys. Rev. D \textbf{16} 3103 (1977).

\bibitem{lsnd_xsec}
L.B. Auerbach \textit{et al.}, 
Phys. Rev. C \textbf{66} 015501 (2002).

\bibitem{cowan}
G. Cowan, 
\textit{Statistical Data Analysis} (Clarendon, Oxford, 1998). 

\bibitem{roe}
B. Roe, 
\textit{Probability and Statistics in Experimental Physics} (Springer, New York, 2001). 

\bibitem{jsns2_tdr}
S. Ajimura \textit{et al.},
arXiv:1705.08629.



\end{thebibliography}
\end{document}